\documentclass{article}
\usepackage{ijcai22}

\usepackage{times}
\usepackage{soul}
\usepackage{url}
\usepackage[hidelinks]{hyperref}
\usepackage[utf8]{inputenc}
\usepackage[small]{caption}
\usepackage{graphicx}
\usepackage{amsmath}
\usepackage{amsthm}
\usepackage{booktabs}
\usepackage{algorithm}
\usepackage{algorithmic}
\usepackage{xspace}
\usepackage[dvipsnames]{xcolor}
\usepackage{booktabs}
\usepackage{multirow}
\usepackage{amsfonts}

\newcommand{\nc}{\newcommand}
\nc{\ale}[1]{\noindent{\color{blue}\textbf{Alessandro:} #1}}
\nc{\tdn}[1]{\noindent{\color{Bittersweet}\textbf{Thomas} #1}}

\nc{\calA}{{\ensuremath{\cal{A}}}} \nc{\calB}{{\ensuremath{\cal{B}}}}
\nc{\calC}{{\ensuremath{\cal{C}}}} \nc{\calD}{{\ensuremath{\cal{D}}}}
\nc{\calE}{{\ensuremath{\cal{E}}}} \nc{\calF}{{\ensuremath{\cal{F}}}}
\nc{\calG}{{\ensuremath{\cal{G}}}} \nc{\calH}{{\ensuremath{\cal{H}}}}
\nc{\calI}{{\ensuremath{\cal{I}}}} \nc{\calJ}{{\ensuremath{\cal{J}}}}
\nc{\calK}{{\ensuremath{\cal{K}}}} \nc{\calL}{{\ensuremath{\cal{L}}}}
\nc{\calM}{{\ensuremath{\cal{M}}}} \nc{\calN}{{\ensuremath{\cal{N}}}}
\nc{\calO}{{\ensuremath{\cal{O}}}} \nc{\calP}{{\ensuremath{\cal{P}}}}
\nc{\calQ}{{\ensuremath{\cal{Q}}}} \nc{\calR}{{\ensuremath{\cal{R}}}}
\nc{\calS}{{\ensuremath{\cal{S}}}} \nc{\calT}{{\ensuremath{\cal{T}}}}
\nc{\calU}{{\ensuremath{\cal{U}}}} \nc{\calV}{{\ensuremath{\cal{V}}}}
\nc{\calW}{{\ensuremath{\cal{W}}}} \nc{\calX}{{\ensuremath{\cal{X}}}}
\nc{\calY}{{\ensuremath{\cal{Y}}}} \nc{\calZ}{{\ensuremath{\cal{Z}}}}
\nc{\calz}{{\ensuremath{\cal{z}}}}

\nc{\bmnull}{{\ensuremath{\boldsymbol{0}}}} 
\nc{\bma}{{\ensuremath{\boldsymbol{a}}}} 
\nc{\bmb}{{\ensuremath{\boldsymbol{b}}}}
\nc{\bmc}{{\ensuremath{\boldsymbol{c}}}} \nc{\bmd}{{\ensuremath{\boldsymbol{d}}}}
\nc{\bme}{{\ensuremath{\boldsymbol{e}}}} \nc{\bmf}{{\ensuremath{\boldsymbol{f}}}}
\nc{\bmg}{{\ensuremath{\boldsymbol{g}}}} \nc{\bmh}{{\ensuremath{\boldsymbol{h}}}}
\nc{\bmi}{{\ensuremath{\boldsymbol{i}}}} \nc{\bmj}{{\ensuremath{\boldsymbol{j}}}}
\nc{\bmk}{{\ensuremath{\boldsymbol{k}}}} \nc{\bml}{{\ensuremath{\boldsymbol{l}}}}
\nc{\bmm}{{\ensuremath{\boldsymbol{m}}}} \nc{\bmn}{{\ensuremath{\boldsymbol{n}}}}
\nc{\bmo}{{\ensuremath{\boldsymbol{o}}}} \nc{\bmp}{{\ensuremath{\boldsymbol{p}}}}
\nc{\bmq}{{\ensuremath{\boldsymbol{q}}}} \nc{\bmr}{{\ensuremath{\boldsymbol{r}}}}
\nc{\bms}{{\ensuremath{\boldsymbol{s}}}} \nc{\bmt}{{\ensuremath{\boldsymbol{t}}}}
\nc{\bmu}{{\ensuremath{\boldsymbol{u}}}} \nc{\bmv}{{\ensuremath{\boldsymbol{v}}}}
\nc{\bmw}{{\ensuremath{\boldsymbol{w}}}} \nc{\bmx}{{\ensuremath{\boldsymbol{x}}}}
\nc{\bmy}{{\ensuremath{\boldsymbol{y}}}} \nc{\bmz}{{\ensuremath{\boldsymbol{z}}}}

\title{Graph Neural Networks for Microbial Genome Recovery}

\author{
Andre Lamurias$^1$
\and
Alessandro Tibo$^1$\and
Katja Hose$^1$\and
Mads Albertsen$^2$\And
Thomas Dyhre Nielsen$^1$
\affiliations
$^1$Department of Computer Science, Aalborg University, Aalborg, Denmark\\
$^2$Center for microbial communities, Aalborg University, Denmark\\
\emails
\{andrel,alessandro,khose,tdn\}@cs.aau.dk,
ma@bio.aau.dk,
}

\newcommand{\backup}[1]{}
\newcommand{\binner}{\textsc{VaeG-Bin}\xspace}

\begin{document}

\maketitle

\begin{abstract}

  Microbes have a profound impact on our health and environment, but our understanding of the diversity and function of microbial communities is severely limited. Through DNA sequencing of microbial communities (metagenomics), DNA fragments (reads) of the individual microbes can be obtained, which through assembly graphs can be combined into long contiguous DNA sequences (contigs). Given the complexity of microbial communities, single contig microbial genomes are rarely obtained. Instead, contigs are eventually clustered into bins, with each bin ideally making up a full genome. This process is referred to as metagenomic binning. 
  
  Current state-of-the-art techniques for metagenomic binning rely only on the local features for the individual contigs. These techniques therefore fail to exploit the similarities between contigs as encoded by the assembly graph, in which the contigs are organized.
  In this paper, we propose to use Graph Neural Networks (GNNs) to leverage the assembly graph when learning contig representations for metagenomic binning.
  Our method, \binner, combines variational autoencoders for learning latent representations of the individual contigs, with GNNs for refining these representations by taking into account the neighborhood structure of the contigs in the assembly graph. 
  We explore several types of GNNs and demonstrate that \binner recovers more high-quality genomes than other state-of-the-art binners on both simulated and real-world datasets.
  
\end{abstract}

\section{Introduction}

Microbial communities have a direct impact on human health and our environment and they play an essential role in achieving the sustainable development goals~\cite{akinsemolu2018role,timmis2017contribution}, in particular \emph{good health and well-being} (SDG-3), \emph{life below water} (SDG-14), and \emph{life on land} (SDG-15), to name a few. Being able to explore the microbial potential for the general good does, however, require an astute understanding of the microbial world in terms of, among others, diversity and function. Metagenomics studies microbial communities at the DNA level, and in theory it is possible to recover the genomes of all the microbes in a sample. However, this is a complex task since DNA sequencing technologies can only produce fragments of the full genome, and, due to the incompleteness of current reference databases, the full genome of most microbes in environmental samples remains unknown~\cite{pasolli2019extensive}.

The process of recovering genomes from the fragmented sequencing data is called binning.
In general, binning is a two-step process, where the first step defines a notion of similarity between DNA sequences and the second step consists of grouping these sequences into clusters, which are referred to as bins.
The input to the binning process is a set of assembled contiguous DNA sequences (contigs). Contigs are obtained by representing the fragmented sequences as a graph, called an assembly graph, where each node represents a contig and the edges represent overlaps between contigs.
Most binners~\cite{yang2021review} only use local features of the individual contigs, thus failing to take full advantage of the relational information embedded within the assembly graph.
Since, by construction, connected contigs share similar DNA sub-fragments, we hypothesize that the assembly graph holds potentially important information that can be exploited during the binning process. 

With the recent successes of applying deep neural networks to various problems, there has also been an increasing focus on adapting such approaches to graph data structures.
Graph Neural Networks (GNNs) take advantage of the connectivity information in a graph and can be used to perform node, edge, and graph-level tasks.
Several types of GNNs have been proposed, such as Graph Convolutional Networks (GCN)~\cite{kipf2017semi}, GraphSAGE~\cite{hamilton2017inductive}, and Graph Attention Networks~\cite{VelickovicCCRLB18}.
Concurrently with the present work, GNNs have also been used for metagenomics binning, showing promising results~\cite{xue2021repbin,lamurias2022metagenomic}.

In this paper, we present \binner, a binning approach based on Graph Neural Networks (GNN), integrating local features obtained through a Variational Autoencoder (VAE)~\cite{KingmaW13} with global features learned from the assembly graph.
We compare \binner to existing state-of-the-art binning techniques on real-world and simulated datasets and demonstrate a significant improvement compared to state of the art using standard genome-recovery evaluation metrics.
The code and data used in the experiments will be made available upon acceptance.

\begin{figure*}[!t]
    \centering
    \includegraphics[width=0.99\textwidth]{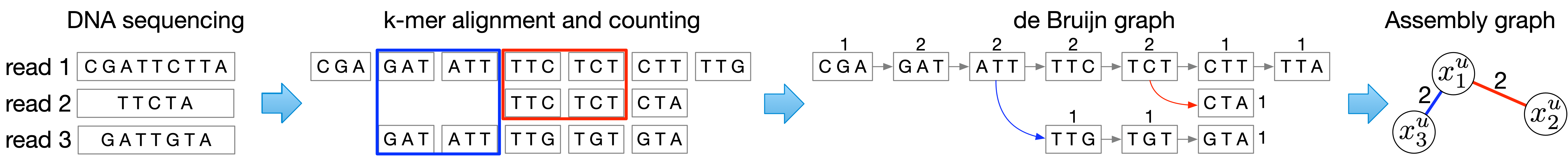}
    \caption{Assembly graph generation. The DNA sequences are read from an environmental sample, converting the raw signal to one of four bases. While finding the best alignment, the reads are broken into k-mers (k$=3$ in this example, but usually much larger), and matching k-mers are aligned. The overlapping k-mers are aggregated and organized in a de Brujin graph. Here, each path from the root (\texttt{C G A}) to an end node ((\texttt{T T A}), (\texttt{C T A}), (\texttt{G T A})) generates a contig. For example, the sequence \texttt{C G A T T G T A} is a contig. The integer numbers reported in the de Brujin graph corresponds to the number of times k-mers overlap for different readings. Finally, each contig is associated with a node in the assembly graph. The edge weights are the fraction of reads that overlap at the intersection of the contig pair, in this case, two reads align to both edges.}
    \label{fig:assembly}
\end{figure*}

\section{Domain background} 
The genome of an organism is the collection of all its genetic information, represented in the form of a sequence of DNA bases.
In an environmental sample, we encounter a combination of genomes from multiple individuals.
The general metagenomic workflow starts then by extracting and sequencing DNA fragments from an environmental sample.
High-throughput sequencing produces a raw electrical signal that is then converted into a sequence consisting of the four DNA bases (ATCG).
This procedure generates millions to billions of reads, which may originate from any of the genomes of all the organisms in the sample.
Reads can have variable lengths, and depending on the technology used, they are classified as short reads (100-150 bases) or long reads (2-30k bases).
While longer read lengths are preferable to fully reconstruct the genome, up until recently long reads were also more prone to errors~\cite{sereika2021oxford}.

To obtain full microbial genomes, which are in the order of millions of bases, we need to combine these reads into longer sequences.
As one sample may contain numerous identical copies of a microbial species, the reads will be a collection from these organisms starting at random points of the genome, and hence have partial overlaps if enough reads are sampled.
The process of combining these reads is called assembly and it involves finding overlaps between reads to obtain contiguous sequences, called contigs. Specifically, reads (through k-mers) are encoded in a de Brujin graph~\cite{compeau2011apply} that serves as a generator, where each walk of the de Brujin graph corresponds to a contig. By finding sub-sequence overlaps (k-mers) within the reads, an assembly graph is generated, where each node corresponds to a contig and an edge represents a possible continuation of that contig in the genome.
The number of reads that overlap on the same position is called coverage or depth. Figure~\ref{fig:assembly} shows an example assembly graph generation starting from the reads.

Since the genome of each organism will be split into several contigs, advanced methods are required to recover high-quality genomes from a set of contigs.
These methods are referred to as binners since they partition contigs into different bins.
As reads correspond to actual DNA sequences present in the sample, the read coverage of a contig will be correlated to the number of organisms in the sample.
This property is called abundance and is a useful feature to bin contigs since contigs from the same genome should have similar abundances~\cite{albertsen2013genome}.
Another useful property is the k-mer frequencies of a contig, generally of size 3 or 4, which should also be similar for contigs from the same genome (also known as k-mer composition)~\cite{burge1992over}.
An important set of genes are the Single Copy Genes (SCG), which occur only once in the full genome but which are essential for the functioning and reproduction of the microbes. Information about the single copy genes can be incorporated into the binning process, since two contigs with the same SCG must belong to different genomes and should therefore appear in different bins.
Therefore, the aim of the binning task is to partition contigs into bins that contain a single copy of all the genes in the set of SCGs.

\section{Related Work}
In recent years, several binners have been proposed based on k-mer composition and abundance features~\cite{yang2021review}. 
One of the best-performing binners based on these features is MetaBAT2~\cite{kang2019metabat}.
MetaBAT2 uses these two features to compute a pairwise distance matrix for all contig pairs, calculated with a k-mer frequency distance probability and abundance distance probability.
The former is based on an empirical posterior probability obtained from a set of reference genomes.
MaxBin2~\cite{wu2016maxbin} is another method that uses an Expectation-Maximization algorithm to estimate the probability of a contig belonging to a particular bin.
The SCGs associated  with each contig are used to estimate the number of bins.
Although more k-mer composition and abundance methods have been proposed~\cite{lu2017cocacola,yu2018bmc3c}, MetaBAT2 and MaxBin2 are the most established and commonly used ones.

More recently, deep learning-based methods have been used to improve metagenomic binning.
Deep learning models present an advantage over other statistic methods since these types of models have the potential to learn complex patterns in the data that would be difficult to model with other methods. 
VAMB~\cite{nissen2021improved} is a binner based on a variational autoencoder that encodes k-mer composition and abundance features in a low dimensional embedding that can lead to improved binning results.
However the usage of deep learning for metagenomics is still in its early stages and very few works have explored how to adapt existing algorithms for these problems, in particular for the most recent sequencing technologies that produce longer reads~\cite{sereika2021oxford}.

Some recent works have attempted to use the assembly graph to improve metagenomic binning.
The common assumption is that contigs that are linked in the assembly graph should also be binned together.
For example, GraphBin~\cite{mallawaarachchi2020graphbin} refines bins from other tools using information from the assembly graph. Specifically, GraphBin navigates the assembly graph using a label propagation algorithm and refines clusters that were separated in the binning process, but which nevertheless contain contigs that are linked in the assembly graph.
However, GraphBin uses the assembly graph only as a post-processing step, and does not integrate it into the full binning process. By relying on the assembly graph only as a last step of the binning process, errors can potentially be introduced if the relational structure in the assembly graph is not carefully used, e.g., contigs may be incorrectly assigned to bins due to misleading or erroneous links in the assembly graph.
This is more likely to occur in complex samples, where variants of the same species (strains) exist, thereby making it more likely that the assembly graph contains links between contigs even if these contigs belong to different genomes.




\section{Methodology}
In the following, we denote with $x$ vectors in $\mathbb{R}^n$ (including scalars) and $\calX$ for sets. In \binner, the data is always represented as an assembly graph $G=(\calV, \calE)$, where $\calV$ and $\calE$ represent the sets of nodes and edges, respectively. Each node $u \in \calV$ is a sequence of length $\ell(u) \in \mathbb{N}$, but it is represented as a tuple of features $x^u=(x^u_t \in \mathbb{R}^{n_t}, x^u_a \in \mathbb{R}^{n_a} )$, where $x^u_t$ represents the k-mer frequencies, and $x^u_a$ represents the relative abundances. In all our experiments, we consider $x^u$ as the concatenation of $x^u_t$ and $x^u_a$ which has size $n_t + n_a$. The dimensionalities $n_t$ and $n_a$ of both vectors depend on the specific datasets. Each node $u \in \calV$ is either associated with a genome (categorical) label $y^u$ or a set of SCGs $\hat{\calY}(u)$ (up to 104)  when genome labels are not available. The SCGs are predicted by CheckM~\cite{parks2015checkm}, a standard metagenomic evaluation tool. Note that in both scenarios \binner remains completely unsupervised with respect to the genome labels, which are only used in the quantitative evaluations. In contrast to classical graph problems, the set of edges in the assembly graph may contain several false positives. To mitigate this issue, each edge $(u,v) \in \calE$ is assigned a weight $w(u,v) \in [0,1]$, which represents the fraction of reads that overlap with both nodes of that edge and can thus be seen as an edge confidence. Here, $0$ and $1$ mean low and high confidences, respectively.
\begin{figure}
    \centering
    \includegraphics[width=0.95\columnwidth]{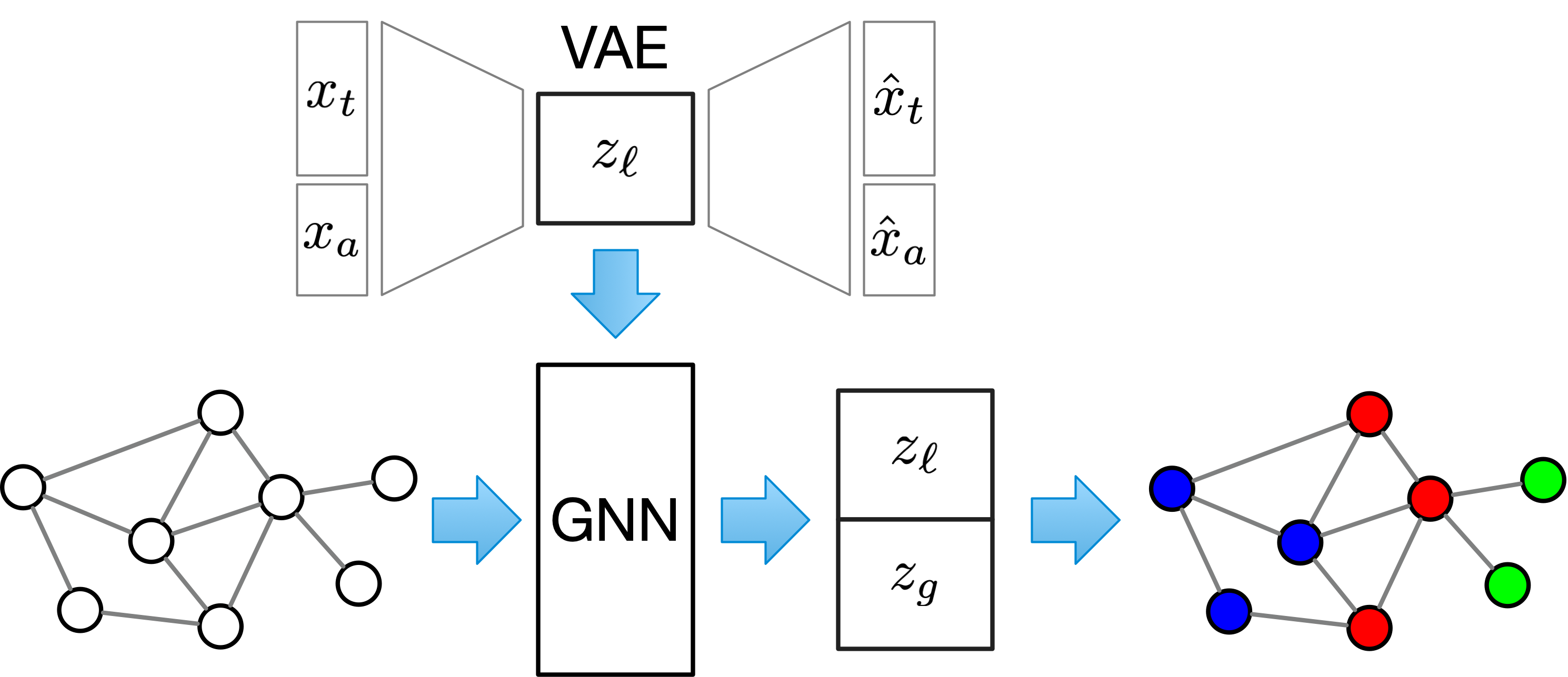}
    \caption{A Variational Autoencoder (top) is used to learn node representations $z_{\ell}$. The graph structure and $z_{\ell}$ are fed into a graph neural network (bottom) which outputs features $z_g$ depending on the graph structure. Finally, $z_{\ell}$ and $z_g$ are concatenated and clustered.}
    \label{fig:framework}
\end{figure}


The \binner framework, depicted in Figure~\ref{fig:framework}, consists of a local and a global feature extractor for the nodes in $\calV$. The local features (contig-specific representations) $z_{\ell}$ are learned with a Variational Autoencoder (VAE), while we adopt a graph neural network (GNN) approach for learning global features (graph representations). The GNN takes as input $z_{\ell}$ and $G$ and produces a global representation for each node, $z_g$. Finally, $z_{\ell}$ and $z_g$ are concatenated and fed into a clustering algorithm to discover the bins. In the following sections, the terms clusters and bins are used interchangeably. 
Recall that we aim at determining the clusters assignments in which each cluster contains as many unique SCGs nodes as possible. While our approach remains completely unsupervised, our aim is reflected in Equation~\ref{eq:gnn:loss} as described below.

\subsection{Contig-specific representations}
We first generate contig-specific representations by encoding k-mers $x_t$ and relative abundances $x_a$ with a VAE (see Figure~\ref{fig:framework}). A VAE consists of an encoder $E$, paramaterized by $\theta_E$ and a decoder $D$, parameterized by $\theta_D$. Each $x_t$ is normalized to have zero mean and unitary variance, while each component of $x_a$ is normalized to have a sum equal to $1$ across all the relative abundances. The loss function used to train the VAE is adopted from \cite{nissen2021improved} and consists of three components\footnote{In all of our experiments, $w_a=(1-\alpha)\log(n_a + 1)^{-1}$, $w_t = \alpha/n_t$, and $w_{kl}=(n_z\beta)^{-1}$, where $n_z$ is the dimension of $\mu_z$, $\alpha=0.15$ and $\beta=200$. See also \cite{nissen2021improved}.}    :
\begin{equation*}
\begin{split}
    J(x_t,x_a; \theta_E, \theta_D)  = \ & w_a \ x_a^T \log (\hat{x}_a + \epsilon) \\
    & + w_t \ \|x_t - \hat{x}_t \|^2 \\
    & -  w_{kl} \ D_{KL}(\calN(\mu_z, \sigma_z) || \calN(0, I)),  \nonumber
\end{split}
\end{equation*}where $D_{KL}$ is Kullback-Leibler divergence, $\epsilon$ is a small constant, $(\mu_z$, $\log \sigma^2)$ = $E((x_t, x_a);\theta_E)$, and $(\hat{x}_t, \hat{x}_a)$ = $D(z; \theta_E)$. Thus, the reconstruction error is separated into two terms capturing the k-mer compositions and abundances of the contigs, respectively. Note that here $z$ is sampled by using the reparametrization trick on $\mu_z$ and $\log \sigma^2$. Finally, we use $z_{\ell} = \mu_z$ produced by $E$ as node features in the following sections; in preliminary experiments we found that $\sigma$ attains very small values and is therefore not included in the feature representation.


\subsection{Graph representations}
A GNN enables learning of node features that depend on the node neighborhoods. In particular, GNNs aggregate the neighbors' information through the following generic graph convolutional layer:
\begin{equation}\label{eq:gnn:generic}
    z^u_g = \alpha_{u,u} \Theta_1  z^u_{\ell} + \Theta_2\sum_{ v \in \calN(u)} \alpha_{u,v} z^v_{\ell},
\end{equation}where $z^u_{\ell}$ and $z^v_{\ell}$ are the feature vectors produced by the VAE associated with nodes $u$ and $v$, respectively. $\Theta_1$ and $\Theta_2$ are learnable parameterized matrices and $\alpha_{u,v} \in \mathbb{R}$ is a scalar for weighting the contribution of each node in the neighborhood. Note that multiple layers, as defined in Equation~\ref{eq:gnn:generic}, can be stacked together in order to provide representations that depend on nodes at larger depths in the graph. Finally, each graph convolutional layer can also be intermixed with standard neural network layers. 

We remark that our framework, \binner, is generic with respect to the GNN. In our experiments (see Section~\ref{sec:experiments}) we have evaluated \binner on three classical GNN architectures: GCN~\cite{kipf2017semi}, GraphSAGE~\cite{hamilton2017inductive}, and GAT~\cite{VelickovicCCRLB18}.

The key to \binner is the loss function used
to train the GNN, defined on pairs of GNN outputs.
\begin{eqnarray}\label{eq:gnn:loss}
    J(z^u_g, z^v_g ; \Theta) & = & w(u,v) \log (\sigma (<{z_g^u}, z_g^v>)) \nonumber \\
    & + & (1 - w(u,v)) \log (1-\sigma (<{z_g^u}, z_g^v>)) \nonumber \\
    & + & \mathbb{I}[ | \hat{\calY}(u) \cap \hat{\calY}(v) | > 0] e^{-\|z_g^u-z_g^v\|^2},
\end{eqnarray}where $\Theta$ are the GNN parameters, $\sigma$ is the sigmoid function, $<\cdot,\cdot>$ denotes the scalar product, and $\mathbb{I}$ is the indicator function. The first two terms of the loss represent the weighted binary cross-entropy between connected and disconnected nodes in the assembly graph. The last term in the loss encourages different features for nodes with the same SCGs.
For the sake of simplicity, we consider all the edges with unitary weights. For GCNs, Equation~\ref{eq:gnn:generic} becomes:
\begin{equation*}
    z_g^u = \frac{1}{d_u} \Theta  z_{\ell}^u + \Theta \sum_{ v \in \calN(u)} \frac{1}{\sqrt{d_u d_v}} z_{\ell}^v,
\end{equation*}where $d_u = 1 + |\calN(u)|$, and $\Theta=\Theta_1=\Theta_2$.
For GraphSAGE, Equation~\ref{eq:gnn:generic} takes the form:
\begin{equation*}
    z_{g}^u = \Theta_1 z_{\ell}^u + \Theta_2 \frac{1}{|\calN(u)|}\sum_{ v \in \calN(u)}  z_{\ell}^v.
\end{equation*}Note that in our experiment, following~\cite{hamilton2017inductive}, we also aggregate neighborhoods with LSTMs. In Section~\ref{sec:experiments} we denote with \textsc{GraphSAGE-M} and \textsc{GraphSAGE-L} the versions that use average and LSTM aggregations, respectively.
For GATs, Equation~\ref{eq:gnn:generic} is specified as:
\begin{equation*}
    z_g^u = \alpha_{u,u} \Theta  z_{\ell}^u + \Theta \sum_{ v \in \calN(u)}  \alpha_{u,v} z_{\ell}^v, 
\end{equation*}where
\begin{equation*}
    \alpha_{u,v} = \frac{
      \exp ( \textsc{L-ReLU}( a^T (\Theta z_{\ell}^u || \Theta z_{\ell}^v)) )
    }
    {
      \sum\limits_{k \in \calN(u) \cup \{u \}}\exp ( \textsc{L-ReLU}( a^T (\Theta z_{\ell}^u || \Theta z_{\ell}^k)) )
    },
\end{equation*}with $a$ being a learnable parameter and \textsc{L-ReLU} the leaky ReLU activation function.

\subsection{Clustering and evaluation}
\label{sec:clustering_and_evaluation}
For the sake of consistency, we adopt
the same cluster algorithm used in~\cite{nissen2021improved}, a modified version of the $k$-medoids algorithm, which does not require an initial number of clusters.
The clustering algorithm receives as input the concatenation of the contig-specific and graph representations, i.e., $z^u=(z^u_{\ell}, z^u_g)$. 
This algorithm consists of a three-step process: it first finds a seed medoid by picking a random $z^u$ associated with a node and calculates the cosine distance to all other $z^v$. 
If any node has more neighbors than the current medoid within a small radius, that one is picked as the new medoid.
The second step consists in determining the cluster radius. The distance from the chosen medoid to all other nodes is calculated, and the algorithm tries to find an optimal distance threshold that includes most of the nearby nodes, but small enough to exclude distant nodes, which should correspond to a local minimum in a histogram plot of the distances.
The third step consists in removing the nodes within that threshold from the list of nodes to cluster and returning to step one until no more unclustered nodes are left. A more detailed description of the algorithm can be found in \cite{nissen2021improved}.

To evaluate the quality of the bins (clusters), we adopted the completeness (see Equation~\ref{eq:compl}) and the contamination (see Equation~\ref{eq:cont}) criteria. Both criteria are domain-specific and indicate the quality of the cluster, according to the Minimum Information about a Metagenome-Assembled Genome (MIMAG) standard set by the Genomic Standards Consortium~\cite{bowers2017minimum}.
Completeness indicates whether the genome is suitable for a specific downstream analysis, while contamination indicates the fraction of the genome that might be contaminated with sequences from other genomes.
These two metrics are required to submit a genome to public databases and to report it in publications.
Using these criteria, we can classify a bin as High Quality (HQ) if completeness $>$ 0.9  and contamination $<$ 0.05, and as Medium Quality (MQ) if completeness $>$ 0.5  and contamination $<$ 0.1\footnote{HQ bins are also required to have the 5S, 16S and 23S rRNA genes and 18 tRNA genes, however, we did not check for these properties in this work.}.

The recommended way of calculating these metrics is to use the list of SCGs as ground truth (recall that these genes are present only once in the genomes of nearly all bacteria).
Some SCGs are collocated, meaning that they are in close proximity in the DNA, and so their occurrences are not fully independent.
For this reason, the ground truth is defined in terms of a set of sets of SCGs, ${\calG}_M$, where each set of SCGs represents a group of collocated SCGs.

The completeness of a bin is given by:
\begin{equation}\label{eq:compl}
    \textsc{Comp}({\calG}_M, \hat{\calY} ) = \frac{1}{|{\calG}_M|}\sum_{\calG \in {\calG}_M} \frac{|\calG \cap \hat{\calY}|}{|\calG|},
\end{equation}where $\hat{\calY}$ represents the multiset of SCGs associated with the nodes of a single bin. The completeness takes value 1 (the maximum) when all genes from $\calG_M$ are identified in the bin. Completeness can be associated with the concept of recall, since it measures the fraction of retrieved genes in the bins.

The contamination of a bin is defined as
\begin{equation}\label{eq:cont}
 \textsc{Cont}({\calG}_M, \hat{\calY}) = \frac{1}{|{\calG}_M|} \sum_{\calG \in {\calG}_M}  \frac{1}{|\calG|} \big(\sum_{g \in \calG} \big(\sum_{y \in \hat{\calY}} \mathbb{I}[g=y] \big) - 1\big),
\end{equation}where $\mathbb{I}$ is the indicator function which is 1 if $g$ is equal to $y$, and 0 otherwise. Here, we assume that if $g \notin \hat{\calY}$ the inner most summation in Equation~\ref{eq:cont} is 0. 
There is no maximum value of contamination, since it will depend on the number of times an SCG is duplicated, i.e., a value of 1 means that on average all genes from $\calG_M$ are duplicated once, and 2 means that all genes have two additional copies on average.

For simulated datasets, the genomes in the dataset are known. Therefore, it is possible to map the node sequences to those genomes and obtain the ground truth genome label $y^u$ of each node.
We followed the evaluation criteria for simulated datasets with ground truth labels as described in~\cite{meyer2018amber}: using the AMBER evaluation tool, we evaluate precision and recall of each bin according to the labels of the nodes that constitute the cluster.
If a bin contains all the nodes associated with one label, then that bin will have a recall of 1, and if it does not contain nodes of any other labels, it will have a precision of 1.
In these metrics, we also take into account the length of the nodes, because longer nodes will have a bigger impact on recovering the genome sequence than smaller nodes.

Average precision (AP), average recall (AR), and F1 are thus defined as follows:
\begin{gather}
    \textsc{AP}  =  \frac{1}{K}\sum_{k=1}^{K}\frac{TP_k}{TP_k+FP_k} 
     \quad \textsc{AR}  =  \frac{1}{K}\sum_{k=1}^{K}\frac{TP_k}{TP_k+FN_k} \nonumber \\
    \textsc{F1}  = \frac{2 \cdot \textsc{AP} \cdot \textsc{AR}}{\textsc{AP}+\textsc{AR}}, \nonumber
\end{gather}where $K$ is the number of clusters and
\begin{gather}
   TP_k  =  \sum_{u \in C_k} \ell(u) \mathbb{I}[y^k=y^u]
   \quad FP_k  =  \sum_{u \in C_k} \ell(u) \mathbb{I}[y^k\neq y^u] \nonumber \\
   FN_k  =  \sum_{u \notin C_k} \ell(u) \mathbb{I}[y^k= y^u]. \nonumber
\end{gather}
Here, $y^k$ is the label associated with the cluster $C_k$, calculated as the majority label of the node labels belonging to $C_k$. Similar to the previous criterion, we considered as HQ bins those with $>0.9$ recall and $>0.95$ precision, and as MQ bins those with $>0.5$ recall and $>0.9$ precision.

\section{Experiments}\label{sec:experiments}

\begin{table}[!t]
  \caption{Datasets used in the experiments. \textsc{Strong100} is a simulated dataset, while the others are real-world datasets. $n_t$ is the dimension of the k-mer frequency features and $n_a$ is the dimension of the abundance features.}
  \label{tab:datasets}
  \setlength\tabcolsep{3.0pt}
  \centering
  \begin{small}
    \begin{sc}
      \begin{tabular}{lcccc}
        \toprule
        Datasets  & \# Nodes & \# Edges & $n_t$ & $n_a$
        \\ 
        \midrule
        Strong100 & 852       & 1,952     & 136 & 1 \\
        \midrule
        AalE      & 45,831    & 33,173    & 136 & 4 \\
        Mari      & 41,559    & 35,001    & 136 & 4 \\
        DamH      & 38,578    & 34,186    & 136 & 4 \\          
        \bottomrule
      \end{tabular}
    \end{sc}
  \end{small}
\end{table}

\paragraph{Data}
We perform experiments on one simulated dataset and three Wastewater Treatment Plant (WWTP) datasets (Table~\ref{tab:datasets}). 
Since the benchmark simulated datasets used by other binners do not include the assembly graph, we simulated a new dataset (Strong100).
The simulated dataset was produced using the badread~\cite{wick2019badread} tool (v0.2.0), where we generated reads according to the methodology proposed in~\cite{quince2020metagenomics}; 
we simulate reads from 100 strains, corresponding to 50 species, with randomly generated abundances.
Badread is a read simulator developed specifically for long-reads, taking into consideration the error rate of these technologies.
In this way, we also generate a dataset that is up-to-date with the current state of DNA sequencing technologies, where longer reads can be obtained, leading to longer contigs, as well.
We then assemble the contigs with the metaflye~\cite{kolmogorov2020metaflye} tool (v2.9) and ran other binners for comparison.
The WWTP datasets come from a previous study~\cite{singleton2021connecting}.
For the WWTP datasets, we have access to four samples for each WWTP. Recall, that each WWTP is associated with a set of contigs. Therefore, for each contig, the abundance values are stored in the entries of a vector of size four (one for each sample).
While the simulated dataset has ground truth labels, mapping each node to a specific genome, for the real-world datasets we do not have access to this information and we instead follow common practice and estimate the quality of the binning results in terms of the number of high and medium quality bins (see Section~\ref{sec:clustering_and_evaluation}).
The details of the graphs of each dataset are reported in Table~\ref{tab:datasets}.

\paragraph{Parameters}
The input dimensions of each dataset are specified in Table~\ref{tab:datasets}.
The $n_t$ value is the same for all datasets as we used k-mers of size 4 and aggregated k-mers that were the same as their reverse complement.
Both the encoder and decoder of the VAE consist of two hidden layers with 512 nodes and leaky ReLU activations. $\mu_z$ and $\log \sigma^2_z$ have size 32 for the simulated and 64 for the real-world datasets. The VAE are trained by using gradient descent for 500 epochs with a learning rate of $1e^{-3}$.
We use GNNs  with three graph convolutional layers for the real-world datasets and one graph convolutional layer for the simulated dataset. In both cases, the hidden layers consist of 128 nodes and the output $z^u$ has 64 nodes. The learning rate was set to $1e^{-2}$ and we performed 500 epochs of training.

\subsection{Results}
We compare the results of \binner with four competitors on the same datasets, using the default values specified in the corresponding papers.
All the methods take as input the contig sequences and their abundances.
We compare against MetaBAT2~\cite{kang2019metabat} and MaxBin2~\cite{wu2016maxbin}, which are generally considered state-of-the-art~\cite{yue2020evaluating,vosloo2021evaluating}. We also compare against VAMB~\cite{nissen2021improved} and GraphBin~\cite{mallawaarachchi2020graphbin}, the former because it is the only published binner that uses deep learning methods, and the latter because it also takes the assembly graph as input.
GraphBin runs on top of another binner, so it requires the output of another binner as input.
We used MetaBAT2 as the input to GraphBin because it obtained the highest results of the three other binners we considered. We present the results of the simulated and real-world datasets separately due to the different metrics used. We evaluate each of the four binners as well as \binner with the four considered GNNs. To show the stability of \binner, we ran the experiments ten times.

\subsubsection{Simulated data}
\begin{table}[!t]
 \caption{Results on the simulated dataset. AP and AR denotes the average precision and recall over all bins. The F1 score is calculated by considering the average precision and recall. Finally, HQ and MQ refer to the number of High-quality and Medium-Quality bins.} 
 \label{tab:resultssim}
 \centering
 \begin{small}
    \begin{sc}
     \begin{tabular}{lccccc}
        \toprule
        Model & AP & AR & F1 & HQ & MQ \\ 
        \midrule
        MetaBAT2  & 0.905 & 0.592 & 0.716 & \textbf{26} & \textbf{37} \\ 
        VAMB     & \textbf{0.969} & 0.755 & 0.849 & \textbf{26} & 34 \\ 
        MaxBin2   & 0.818  & 0.765 & 0.791 &  14  &  23 \\
        GraphBin &  0.848 & 0.613 & 0.712 & 23 &  34 \\ 
        \midrule
        GCN      & 0.964 & 0.804 & 0.877 & 25±1 & 32±2 \\ 
        GraphSAGE-M & 0.960 & 0.839 & 0.895 & 24±2 & 31±1\\ 
        GraphSAGE-L & \textbf{0.969} & 0.765 & 0.855 & \textbf{26±1} & 34±2 \\ 
        GAT & 0.950 & \textbf{0.863}  & \textbf{0.904} & 18±3 & 25±4 \\ 
        \bottomrule
     \end{tabular}
    \end{sc}
 \end{small}
\end{table}


Table~\ref{tab:resultssim} shows the results obtained on the simulated dataset, where the metrics are calculated on the ground truth labels of the contigs, using the AMBER evaluation tool~\cite{meyer2018amber}.
These results indicate how the methods work in a scenario where the original genome of each contig is known.
In this scenario, the graph-based methods outperform the established binners on almost all metrics. 
In terms of F1-score, GAT achieves the best balance, obtaining however a low number of HQ and MQ bins.
The GraphSAGE-L variant obtained a higher number of HQ bins, at the expense of a lower F1-score.
While the F1-score takes into account the precision and recall of all bins, the HQ and MQ values exclude the lowest quality bins.
Hence, we can have many bins with low F1-score, without affecting the HQ and MQ values.
Although MetaBAT2 obtained the second lowest F1-score, it had the same number of HQ bins as VAMB and GraphSAGE-L, which is the main quality criterion for metagenomic applications.
For downstream analyses, only the HQ bins can be considered recovered genomes, while the others do not have enough quality to be analyzed, because they are too incomplete or too contaminated.

\subsubsection{Real-world data}
%
As shown in Table~\ref{tab:resultsreal}, we can see that most of the GNNs outperform the other methods in terms of HQ bins recovered.
By combining a VAE with a GNN, we can consistently obtain more HQ bins than all other baseline methods.
In particular, in terms of HQ bins, we outperform both VAMB and MetaBAT2, both of which only rely on local contig features and thus fail to take advantage of the relational contig information embedded within the assembly graph.
In terms of MQ bins, we obtain a higher or comparable number of bins relative to the baselines on two out of the three datasets.
Different instantiations of the GNN model have been explored for all three datasets, with the GCN approach obtaining the largest number of high-quality bins.  The other instantiations obtain similar results on some datasets, but not consistently. We hypothesize that this may partly be due to the loss function not being a good proxy for the quality metrics being used during the evaluation, hence more complex models may fail to bring consistent improvements.

\begin{table}[!t]
  \caption{Results on real-world datasets. HQ and MQ refer to the number of High-quality and Medium-Quality bins.} 
  \label{tab:resultsreal}
  \setlength\tabcolsep{3.0pt}
  \centering
  \begin{small}
    \begin{sc}
      \begin{tabular}{lcccccc}
        \toprule
        \multirow{2}[2]{*}{Model} & \multicolumn{2}{c}{AalE} & \multicolumn{2}{c}{Mari} & \multicolumn{2}{c}{Damh} \\
        \cmidrule(lr){2-3} \cmidrule(lr){4-5} \cmidrule(lr){6-7}
        & HQ & MQ & HQ & MQ & HQ & MQ \\ 
        \midrule
        MetaBAT2  & 53 & 175 & 41 & \textbf{155} & 50 & \textbf{219} \\
        VAMB     & 42 & 160 & 34 & 135          & 31 & 132 \\
        MaxBin2   & 20 & 60  & 20 & 70           & 21 & 82  \\
        GraphBin & 16 & 133 & 21 & 123          & 23 & 176 \\ 
        \midrule
        GCN      & \textbf{55±1} & 175±3 
                 & \textbf{46±1} & 154±3 
                 & \textbf{54±1} & 190±4 \\
        GraphSAGE-M & \textbf{55±0} &  175±1 
                    & 44±1 & 148±2 
                    & 51±1 & 187±2 \\
        GraphSAGE-L & 52±1 & \textbf{184±4} 
                    & \textbf{46±2} & 147±3 
                    & 51±1 & 190±4 \\ 
        GAT         & 53±1 & 174±3
                    & 45±1 & 147±2 
                    & 50±1 & 184±3 \\
        \bottomrule
      \end{tabular}
    \end{sc}
  \end{small}
\end{table}
\section{Conclusion}
This paper reports on interdisciplinary research between data science and bioinformatics, addressing the problem of metagenomic binning of contiguous DNA fragments (contigs). This activity is key for understanding the diversity and function of microbial communities, which have a direct impact on both health and the environment and thus play a vital role in addressing the sustainable development goals. We have proposed \binner, a novel methodology for learning feature representations for contigs, combining local feature representations (obtained through a variational autoencoder) with global features learned using a GNN based on the assembly graph in which the contigs are organized. 

We have compared \binner with other state-of-the-art metagenomic binning methods on both simulated and real-world datasets. We observe that by leveraging the relational information in the assembly graph, we can significantly increase the number of high-quality genomes recovered during the subsequent binning process as compared to the state-of-the-art baseline methods.

This work represents an initial step in the exploration of graph learning methods for metagenomic binning and we believe that there are several promising directions for further work. For instance,  
we plan to refine the clustering step in order to better take into account the distribution of single copy genes over the different clusters. This will involve refining the loss function to promote high completeness and low contamination of the clusters. Additionally, an end-to-end approach that incorporates both representation learning and clustering could bring further improvements to this task. 
We expect that the challenges presented by this task will lead to more solutions that benefit both the Artificial Intelligence field and progress on the SDGs.

\section*{Acknowledgments}

The study was funded by research grants from VILLUM FONDEN (34299, 15510) and the Poul Due Jensen Foundation (Microflora Danica)\vspace*{-12pt}


\appendix

\clearpage

\bibliographystyle{named}
\bibliography{ijcai22-condensed}

\end{document}